\documentclass[aps,prd]{revtex4}
\usepackage{subfigure,graphics} 
\usepackage{flafter} 
\begin{document}

\title{
Higher order terms in an improved heterotic $M$-theory}
\author{Ian G. Moss}
\email{ian.moss@ncl.ac.uk}
\affiliation{School of Mathematics and Statistics, Newcastle University, NE1 7RU, UK}

\date{\today}


\begin{abstract}
Curvature-squared terms are added to a consistent formulation of supergravity on manifolds with
boundary which is meant to represent the low energy limit of strongly coupled heterotic string
theory. These terms are necessary for the cancellation of gravitational anomalies and for
reductions to lower dimensions with broken chiral symmetry. The consequences of anomaly cancellation
when flux and extrinsic curvature terms are taken into account have yet to be fully exploited, but
some implications for flux terms are discussed here.
\end{abstract}
\pacs{PACS number(s): }

\maketitle
\section{introduction}

Some time ago, Horava and Witten \cite{Horava:1995qa,Horava:1996ma} proposed that the low energy
limit of strongly coupled heterotic string theory could be formulated as 11-dimensional
supergravity on a manifold with boundary. This opened up the possibility that matter might exist on
a surface embedded in the 11-dimensional spacetime with supergravity taking care of the
gravitational  interactions.  Although the theory has received less attention recently
than type IIB superstring theory, it nevertheless remains a possible starting point for particle
phenomenology \cite{Witten:1996mz,banks96}.

The original formulation of Horava and Witten contained some serious problems which
limited the range of validy of the 11-dimensional limit. These problems where solved
recently using a new formulation of supergravity on manifolds with boundary 
\cite{Moss:2003bk,Moss:2004ck,Moss:2005zw}. The most serious problem affecting the model was that it
was expressed as a series in the factor $\kappa_{11}{}^{2/3}$ 
multiplying the matter action, which worked
well at leading and next-to-leading order but became ill-defined thereafter. This problem was
resolved by a simple modification to the boundary conditions resulting in a low energy theory which
is supersymmetric to all orders in $\kappa_{11}{}^{2/3}$.

The aim of the present paper is to add curvature-squared terms to the new formulation of
supergravity on manifolds with boundary. 
These terms are necessary for the cancellation of
gravitational anomalies \cite{Horava:1996ma,Lukas:1998ew,Moss:2005zw}, 
and they are important for reductions to
lower dimensions with broken chiral symmetry 
\cite{Witten:1996mz,banks96,Lukas:1998tt,Buchbinder:2003pi,Braun:2006th,Gray:2007qy}. 
Higher order terms should
therefore be present if the theory is truly the low energy limit of the strongly coupled heterotic
string. Curvature-squared terms have been included in the boundary action, for example by Lukas et
al. \cite{Lukas:1998tt}, but they have never been shown to be part of a supersymmetric theory
before.

The methodology adopted will be to construct the boundary conditions and the action of the theory
order by order in derivatives, imposing the local symmetries at each stage. Anomaly cancellation
will be brought about by the Green-Schwarz mechanism \cite{Green:1984sg}, modified to accomodate
boundaries \cite{Horava:1996ma,Moss:2005zw}. The results contain all terms with up to five
derivatives and two fermi fields. A remarkable feature is that the action to this order is uniquely
determined, with only one free parameter $\kappa_{11}$. It seems likely that this determinism in 
the theory will occur at higher orders in the curvatures, leaving no room for free parameters apart
from the gravitational coupling.

Before proceding, it will be helpfull to repeat some of the ingredients of the improved version of
low-energy heterotic $M$-theory described in Ref. \cite{Moss:2004ck}. 
The theory is formulated on a manifold ${\cal M}$ with a boundary consisting of two disconnected
components $\partial{\cal M}_1$ and $\partial{\cal M}_2$ with identical topology. The
eleven-dimensional part of
the action is the conventional action for supergravity, with metric $g_{IJ}$, gravitino $\psi_I$ and
antisymmetric tensor $C_{IJK}$ \cite{cremmer78}. The boundary terms which make the supergravity
action supersymmetric are \cite{luckock89},
\begin{equation}
S_0={1\over  \kappa_{11}^2}\int_{\cal\partial M}\left(
\hat K\mp\frac14\bar\psi_A\Gamma^A\Gamma^B\psi_B\right)dv,
\end{equation}
where $K$ is the extrinsic curvature of the boundary and $A,B,\dots$ denote tangential indices. Hats
denote the standardised subtraction of gravitino terms to make a
supercovariant expression. We
shall take the  upper sign on the
boundary component $\partial {\cal M}_1$ and the lower sign on the boundary component 
$\partial {\cal M}_2$. 

There are additional boundary terms with Yang-Mills multiplets, scaled by a parameter $\epsilon$,
\begin{equation}
S_{YM}=-{\epsilon\over \kappa_{11}^2}\int_{\cal \partial M}dv
\left(\frac14{\rm tr}F^2+
\frac12{\rm tr}\bar\chi\Gamma^AD_A(\hat\Omega^{**})\chi
+\frac14\bar\psi_A\Gamma^{BC}\Gamma^A{\rm tr}{F}^*_{BC}\chi\right),
\label{action1}
\end{equation}
where $F^*=(F+\hat F)/2$ and the connection $\Omega^{**}=(\Omega+\Omega^*)/2$. The original
formulation of Horava and Witten contained an extra `$\chi\chi\chi\psi$' term, but it is not
present in the new version. The formulation given in ref. \cite{Moss:2005zw} was only valid to
order $R$, and our aim here is to extend the theory to include $R^2$ terms and beyond. 

The specification of the theory is completed by boundary
conditions. For the tangential anti-symmetric tensor components,
\begin{equation}
C_{ABC}=\mp \frac{\sqrt{2}}{12}\epsilon\,\left(\omega^Y_{ABC}
\mp\omega^\chi_{ABC}\right).\label{cbc}
\end{equation}
where $\omega^Y$ is the Yang-Mills Chern-Simons form and $\omega^\chi$ is a bilinear gaugino term.
These boundary conditions replace the modified Bianchi identity in the old formulation. A
suggestion along these
lines was made in the original paper of Horava and Witten \cite{Horava:1996ma}. For the gravitino,
\begin{equation}
\Gamma^{AB}\left(P_\pm+\epsilon\Gamma P_\mp \right)\psi_A=
\epsilon J_Y{}^A,\label{gbc}
\end{equation}
where $P_\pm$ are chiral projectors using the outward-going normals, $\Gamma$ is a bilinear gaugino
term and $J_Y$ is the Yang-Mills supercurrent. The resulting theory is supersymmetric {\it to all
orders} in the parameter $\epsilon$, but the gauge
anomalies only vanish if the gauge groups on the boundaries are both $E_8$ and
\begin{equation}
\epsilon={1\over 4\pi}\left({\kappa_{11}\over4\pi}\right)^{2/3}.
\end{equation}
Further details of the anomaly cancellation, and additional Green-Schwarz terms, can be found in
Ref. \cite{Moss:2005zw}.

The gravitational anomaly vanishes if we introduce an extra term into Eq. (\ref{cbc}) involving
the Chern-Simons term $\omega^L$ for local Lorentz transformations. The calculations which follow
can be seen as an attempt to find the supersymmetric completion of the new boundary conditions with
the local Lorentz term. These boundary conditions are sufficient to determine the boundary action. 
Section 2 lays down the general strategy and sets up the derivative expansion scheme. 
Section 3 gives results up to fifth order in derivatives for the boundary terms in the
action and for the boundary conditions. The last part of section 3 considers anomaly cancellation
and discusses the generalisation of the earlier results to all orders in the curvature.
The results are collected together in the conclusion.
 
The conventions used follow Weinberg \cite{Weinberg:2000cr}.The metric signature is $-+\dots+$.
The gamma matrices satisfy $\{\Gamma_I,\Gamma_J\}=2g_{IJ}$ and 
$\Gamma^{I\dots K}=\Gamma^{[I}\dots\Gamma^{K]}$. Eleven dimensional vector
indices are denoted by $I,J,\dots$. The coordinate indices on the boundary are
denoted by $A,B,\dots$, tetrad ones by $\hat A,\hat B,\dots$ and the (outward unit) normal direction
by $N$.
 
\section{Supersymmetry transformations}

Construction of the higher order terms is based on the ingenious method introduced by Bergshoeff et
al.  \cite{Bergshoeff:1988nn,Bergshoeff:1989de}. We combine
the spin connection and gravitino derivatives into a pair $\{\omega^-_{ABC},\psi_{AB}\}$ which is
almost a Yang-Mills multiplet. Adding the higher order terms is similar to adding Yang-Mills
multiplets, which we know how to do. Unfortunately, in 11 dimensions, normal components and flux
terms complicate the simple picture and enhance the technical difficulties.

We shall start from the transformation rules for the graviton multiplet and devise a consistent
derivative expansion scheme.  Then we shall construct quantities which are optimised to make the
best
possible Yang-Mills multiplet. In the next section we construct the boundary terms in the action to
fifth order in derivatives. The following section . 
extends the boundary conditions to fifth order in derivatives and confirms that they are
supersymmetric

We shall use the parameter $\alpha$ to keep track of the order of terms in our derivative expansion.
The order of terms should be preserved
by the sypersymmetry transformations, which are
\begin{eqnarray}
\delta e^{\hat I}{}_J&=&\frac12\bar\eta\Gamma^{\hat I}\psi_{J}\label{varg}\\
\delta\psi_I&=&D_I(\hat\Omega)\eta+
{\sqrt{2}\over 288}\left(\Gamma_I{}^{JKLM}-8\delta_I{}^J\Gamma^{KLM}\right)
\eta\hat G_{JKLM}\label{varpsi}\\
\delta C_{IJK}&=&-{\sqrt{2}\over 8}\bar\eta\Gamma_{[IJ}\psi_{K]}\label{varc}
\end{eqnarray}
where $G$ is the abelian flux tensor. We also require that $C_{ABC}\sim \omega^L_{ABC}$ on the
boundary. The ordering we shall use is,
\begin{eqnarray}
R_{ABCD}=O(\alpha^2)&&\psi_A=O(\alpha)\\
G_{NABC}=O(\alpha^2)&&D_{[A}\psi_{B]}=O(\alpha^2)\\
G_{ABCD}=O(\alpha^4)&&D_{[N}\psi_{B]}=O(\alpha^3)
\end{eqnarray}
Additional tangential derivatives increase the order by one. This expansion scheme is consistent
with the Calabi-Yau reductions found in the literature \cite{Lukas:1998tt}, where the small
parameter is related to the curvature of the Calabi-Yau space.

The first quantity we construct is the gravitino curvature $\psi_{AB}$. We start from the
supersymmetry transformation of the tangential gravitino from Eq. (\ref{varpsi}) to two-fermi
order, which can be written,
\begin{equation}
\delta\psi_A={\cal  D}_A\eta=(D_A+{\cal A}_A)\eta
\end{equation}
where $D_A$ uses  the Levi-Civita connection and ${\cal A}_A$ contains a combination of abelian-flux
terms and gamma matrices. The analogue of the
curvature is defined by
\begin{equation}
[{\cal D}_A,{\cal D}_B]={\cal R}_{AB}.
\end{equation} 
The quantity ${\cal R}_{AB}$ is a tensor which takes values in the gamma-matrix algebra,
\begin{equation}
{\cal R}_{AB}=-\frac14R_{ABIJ}\Gamma^{IJ}+2{\cal D}_{[A}{\cal A}_{B]}.
\end{equation}

The new derivative is used to define the gravitino curvature $\psi_{AB}$,
\begin{eqnarray}
\psi_{NA}&=&2{\cal D}_{[N}\psi_{A]}\\
\psi_{AB}&=&2{\cal D}_{[A}\psi_{B]}-2\Gamma_{[A}\psi_{B]N}.\label{gcdef}
\end{eqnarray}
It may help understand this construction to recall that, in the reduction of 11-dimensional
supergravity, the 10-dimensional gravitino is $\psi_A+\Gamma_A\psi_N/2$. If the normal derivatives
vanish, then $\psi_{AB}$ is the usual 10-dimensional gravitino curvature. The supersymmetry
transformation of the gravitino curvature is
\begin{equation}
\delta\psi_{AB}={\cal R}_{AB}\eta-2\Gamma_{[A}{\cal R}_{B]N}\eta.
\end{equation}
Note that $R_{NABC}$ is very small due to Gauss-Codacci relations, of order $\alpha^5$, and this
variation basically depends on
$R_{ABCD}$ plus abelian-flux terms. 

The supersymmetry transformations are only required on the boundary, where it proves convenient to
decompose the flux-gamma-matrix combinations into tangential and normal components,
\begin{eqnarray}
X={\sqrt{2}\over 72}G_{NABC}\Gamma^{ABC},&
\displaystyle X_A={\sqrt{2}\over 8}G_{NABC}\Gamma^{BC},&\\
Y={\sqrt{2}\over 288}G_{ABCD}\Gamma^{ABCD},&
\displaystyle Y_A={\sqrt{2}\over 24}G_{ABCD}\Gamma^{BCD},&
Y_{AB}={\sqrt{2}\over 8}G_{ABCD}\Gamma^{CD}.\label{ydef}
\end{eqnarray}
For example,
\begin{eqnarray}
{\cal A}_A&=&\Gamma_A(\Gamma_NX+Y)-\Gamma_NX_A-Y_A,\\
{\cal A}_N&=&-2X+\Gamma_NY.
\end{eqnarray}
The supersymmetry parameter is chiral on the boundaries, with $\Gamma_N\eta=\mp\eta$
depending on which boundary we choose. The results below take $\Gamma_N\eta=-\eta$.

Next, we turn to the Levi-Civita spin connection $\omega_{A\hat B\hat C}$. This does not transform
like a Yang-Mills gauge field, but we can adapt an idea from 10 dimensions \cite{Bergshoeff:1989de}
and try the addition of a $G$-flux term,
\begin{equation}
\omega^-_{A\hat B\hat C}= \omega_{A\hat B\hat C}+
{1\over\sqrt{2}}G_{NA\hat B\hat C}.\label{msc}
\end{equation}
The transformation rules for the pair $\hat\omega^-_{A\hat B\hat C}$ and $\psi_{AB}$ become
\begin{eqnarray}
\delta\hat\omega^-_{A\hat B\hat C}&=&
-{1\over 2}\overline\eta\Gamma_A\psi_{\hat B\hat C}+y_{A\hat B\hat C}\label{domega}\\
\delta\psi_{AB}&=&-\frac14R^-{}_{CDAB}\Gamma^{CD}\eta+y_{AB}\eta,\label{dpsiab}
\end{eqnarray}
where the minus superscript on the curvature indicates use of the $\omega^-$ connection. The leading
terms are $O(\alpha^2)$, whereas $y_{A\hat B\hat C}$ and $y_{AB}$ are both $O(\alpha^3)$. These
correction terms 
are given by
\begin{eqnarray}
y_{A\hat B\hat C}&=&
-{1\over 4}\overline\eta\Gamma_{\hat B\hat C}\psi_{NA}
-{1\over 2}\overline\eta\{\Gamma_{\hat B\hat C},X\}\psi_A
+{1\over 2}\overline \eta e_{A\hat B}\psi_{N\hat C}+O(\alpha^5),\\
y_{AB}&=&\Gamma_{AB}Y'-\Gamma_{[A}Y'_{B]}-Y'_{AB}+O(\alpha^5).
\end{eqnarray}
where $Y'$ is shorthand notation for $D_NY$. Note that $R^-_{ABCD}\ne R^-_{CDAB}$ now that the
connection is no longer a metric connection.  

In order to complete the set of transformation rules, we also need
\begin{equation}
\delta\psi_{NA}=y_{NA}\eta,\label{dpsina}
\end{equation}
where
\begin{equation}
y_{NA}=2D_AX+\Gamma_AY'-Y'_A+O(\alpha^5).
\end{equation}
We have not made any modification to the basic supersymmetry rules, and none appears to be necessary
to the order at which we are working. All of the approximations used in this section can be
replaced by exact expressions, but the approximate ones are sufficient for the subsequent sections.

\section{Higher order terms}

\subsection{Boundary terms in the action}

At leading order in $\alpha$, the pair $\{\hat\omega^-_{ABC},\psi_{AB}\}$
form a Yang-Mills multiplet and we
can add this to the boundary conditions and the action in the same way as the existing Yang-Mills
multiplet described in the introduction. We use a new coupling $\epsilon^L$ for the new multiplet,
and anomaly cancellation fixes $\epsilon^L$ \cite{Moss:2005zw},
\begin{equation}
\epsilon^L=-\frac12\epsilon.
\end{equation}
The boundary conditions can be read off Eqs. (\ref{cbc}) and (\ref{gbc}),
\begin{eqnarray}
C_{ABC}&=&-{\sqrt{2}\over 12}\epsilon
\left(\omega^Y_{ABC}+\omega^\chi_{ABC}\right)
+{\sqrt{2}\over 24}\epsilon
\left(\omega^L_{ABC}+\omega^\psi_{ABC}\right),\label{cbc4}\\
\Gamma^{AB}P_+\psi_B&=&\epsilon J_Y{}^A-\frac12\epsilon J_L{}^A,\label{psibc4}\\
\hat K_{AB}-\frac12g_{AB}\hat K&=&\epsilon T^Y_{AB}-\frac12\epsilon T^L_{AB},\label{kbc4}
\end{eqnarray}
where $T^Y_{AB}$ is the Yang-Mills stress tensor and  
\begin{eqnarray}
\omega^\psi_{ABC}&=&\frac14\overline\psi_{DE} \Gamma_{ABC}\psi^{DE},\\
J_L{}^A&=&\frac14\Gamma^{BC}\Gamma^AR_{BCDE}\psi^{DE},\\
T^L_{AB}&=&R^-{}_A{}{}_{CDE}R^-{}_{B}{}^{CDE}
-{1\over 4}g_{AB}R^-{}_{CDEF}R^-{}^{CDEF}+\hbox{$\psi$-terms}.
\end{eqnarray}
The Yang-Mills results imply that Eq. (\ref{kbc4}) is a necessary and sufficient condition for
supersymmetry of the boundary conditions (\ref{cbc4}) and (\ref{psibc4}).

There are also new curvature-squared terms in the boundary action, which we obtain from the
Yang-Mills terms in Eq. (\ref{action1}),
\begin{equation}
S_{RR}={\epsilon\over 2\kappa_{11}^2}\int_{\cal \partial M}dv
\left(\frac14R^-{}_{ABCD}R^-{}^{ABCD}+
\frac12\bar\psi_{BC}\Gamma^AD_A(\omega)\psi^{BC}
+\frac14\bar\psi_A\Gamma^{BC}\Gamma^AR_{BCDE}\psi^{DE}\right).
\label{action2}
\end{equation}
The supersymmetry of the full action with the new boundary terms follows from the gravity-Yang-Mills
calculation given previously \cite{Moss:2004ck}. The modified curvature $R^-_{ABCD}$ has been used
for consistency between the derivative orders of the bosonic and fermionic terms. Note that terms
involving the square of the Ricci tensor only appear at order $\alpha^8$ in the ordering scheme
being used.

Another important property of the full action is that it should be stationary under variations of
the fields about solutions to the field equations with the specified boundary conditions.
Variations of the new boundary term with the tetrad can be decomposed into metric variations and
local Lorenz rotations (see appendix B in \cite{Moss:2004ck}), 
\begin{equation}
\delta S_{RR}={\epsilon\over 2\kappa_{11}^2}\int_{\cal \partial M}dv\left\{
\delta\omega^-_{B\hat C\hat D}\left(
D_AR^{AB\hat C\hat D}-\frac12D_E(\overline\psi_A\Gamma^{BE}\Gamma^A\psi^{\hat C\hat D})
\right)-\frac12\delta g_{AB}T^{LAB}\right\}.
\end{equation}
The surface stress-tensor term is $O(\alpha^4)$ in the derivative expansion  and combines with the
variation of the supergravity action $S_{SG}$  to produce the boundary condition Eq. (\ref{kbc4}).
The metric variation provides a good way to determine the fermion terms in the stress-tensor
\footnote{
Variation of the original supergravity action makes a contribution to $T^L_{AB}$. This was given
incorrectly in section 2 of \cite{Moss:2004ck}. I am grateful to Paul Saffin for pointing out this
mistake.
}.  
The Bianchi
identity and the gravitino field equation together imply that the variation of the surface
connection gives no contribution to $\delta S_{RR}$ at leading order in the derivative expansion.

Variations of the action with the gravitino can be split up in the following way,
\begin{equation}
\delta S_{RR}={\epsilon\over 2\kappa_{11}^2}\int_{\cal \partial M}dv
\left\{\overline{\delta\psi}^{BC}\left(
\Gamma^AD_A\psi_{BC}
+\frac14R_{DEBC}\Gamma^A\Gamma^{DE}\psi_A\right)
+\overline{\delta\psi}_A J_L^A
\right\}.
\end{equation}
The supercurrent term is $O(\alpha^5)$ and contributes to the boundary condition Eq (\ref{psibc4}).
The gravitino field equation can be used to show that the remaining terms are only $O(\alpha^6)$,
and they play no role at leading order.

Finally, it is possible to reduce the 11-dimensional action to 10-dimensions to obtain the low
energy limit of the weakly coupled heterotic superstring. The result of dimensional reduction
agrees with the higher-order action obtained from supergravity 10-dimensions
\cite{Bergshoeff:1988nn}. The curvature-squared  terms obtained from string amplitude calculations
also agree, up to allowed metric redefinitions \cite{Gross:1986mw}. 

\subsection{Boundary conditions and $G$-fluxes}

At fifth order in derivatives new $G$-flux terms begin to contribute to the boundary conditions. Due
to the connection between the boundary conditions and the Green-Schwarz anomaly cancellation
mechanism, these terms allow us to deduce some of the $G$-flux terms in the anomalies. The part of
the supersymmetry transformation which is exactly fifth order in derivatives will be denoted by
$\delta_5$. For the rest of this section we shall drop the Yang-Mills terms.

The fermion boundary condition can be written in the form,
\begin{equation}
P_+\psi_A=-{1\over 24}\epsilon\left(
\Gamma_A{}^{BC}-10\delta_A{}^B\Gamma^C\right)R^-{}_{BCDE}\psi^{DE}
+\epsilon f_A{}^{BC}\psi_{BC},\label{gbc5}
\end{equation}
where $f_A{}^{BC}$ contains $G$-flux terms and gamma-matrices. When we drop the three-fermi terms,
variation of the fermion boundary condition can be done using
Eq. (\ref{dpsiab}) and gamma matrix identities. The fifth order supersymmetry variation
$\delta_5P_+\psi_A$ vanishes for
\begin{equation}
f_A{}^{BC}=\delta_A{}^{[B} Y^{\prime C]}+\frac16\Gamma_AY^{\prime BC},
\end{equation}
where $Y^A$ was defined in Eq. (\ref{ydef}) and prime denotes a normal derivative.

The antisymmetric tensor is a little more complex. The proposed boundary condition is that
\begin{equation}
C_{ABC}={\sqrt{2}\over 24}\epsilon
\left(\omega^L_{ABC}+\omega^\psi_{ABC}+\omega^G_{ABC}\right),\label{cbc5}
\end{equation}
where
\begin{eqnarray}
\omega^L_{ABC}&=&\omega^L_{ABC}(\hat\omega^-)\\
\omega^\psi_{ABC}&=&\frac14\overline\psi_{DE} \Gamma_{ABC}\psi^{DE}
-6\overline\psi_{N[A}\psi_{BC]}-
12\overline\psi_{[AB}X\psi_{C]}\\
\omega^G_{ABC}&=&-\frac{1}{3} *\hat G'_{ABC}{}^{DEF}\hat G_{NDEF}
+3\hat G_{N[A}{}^{DE} \hat G'_{BC]DE}.\label{omegagdef}
\end{eqnarray}
The dual tensor
\begin{equation}
*G'_{ABCDEF}=\frac{1}{24}\varepsilon_{ABCDEF}{}^{PQRS}G'_{PQRS},
\end{equation}
where $G'_{ABCD}=D_NG_{ABCD}=-4D_{[A}G_{BCD]N}$.

Variation of the antisymmetric tensor field on the boundary using the bulk supersymmetry
transformations gives,
\begin{equation}
\delta C_{ABC}=-{\sqrt{2}\over 8}\overline\eta\Gamma_{[AB}\psi_{C]}.
\end{equation}
Since $\eta=P_-\eta$, we can replace $\psi_A$ by $P_+\psi_A$ and use the gravitino boundary conditon
(\ref{gbc5}) to get the fifth order transformation,
\begin{equation}
\delta_5C_{ABC}=-{\sqrt{2}\over 8}\overline\eta\Gamma_{[AB} f_{C]}{}^{DE}\psi_{DE}.
\label{cfp}
\end{equation}
Variation of the terms on the right hand side of the boundary condition using 
Eqs. (\ref{domega}-\ref{dpsina}) gives
\begin{eqnarray}
\delta_5\omega^L_{ABC}&=&
-6\overline\eta R_{[BC}\psi_{NA]}
-12\overline\eta\{R_{[AB},X\}\psi_{C]},\\
\delta_5\omega^\psi_{ABC}&=&
6\overline\eta R_{[BC}\psi_{NA]}
-12\overline\psi_{[AB}XD_{C]}\eta
+12\overline\eta R_{[AB}X\psi_{C]}
+\frac12\overline\eta y^{DE}\Gamma_{ABC}\psi_{DE}
-6\overline\eta y_{NA}\psi_{BC},\\
\delta_5\omega^G_{ABC}&=&-\frac14\sqrt{2}*G'_{ABC}{}^{DEF}\overline\eta\Gamma_F\psi_{DE}
+\frac92\sqrt{2} G'_{AB}{}^{DE}\overline\eta\Gamma_{[C}\psi_{DE]}.\label{dog}
\end{eqnarray}
The best way to deal with the $D_A\eta$ contribution is to remove a total derivative,
\begin{equation}
\delta_5\omega^L+\delta_3\omega^\psi=
-12D_{[A}(\overline\eta X\psi_{BC]})
+12\overline\eta (D_{[A}X)\psi_{BC]}
+\frac12\overline\eta y^{DE}\Gamma_{ABC}\psi_{DE}
-6\overline\eta y_{NA}\psi_{BC},
\end{equation}
where use has been made of the identity
\begin{equation}
D_{[A}\psi_{BC]}=R_{[AB}\psi_{C]}.
\end{equation}
We can absorb the total derivative into an abelian transformation of the $C$ field. After difficult
gamma-matrix manipulations,
\begin{equation}
\delta_5\omega^L+\delta_3\omega^\psi=
-3\overline\eta\Gamma_{[AB}f_{C]}{}^{DE}\psi_{DE}
+\frac14\sqrt{2} *G'_{ABC}{}^{DEF}\overline\eta\Gamma_F\psi_{DE}
-\frac92\sqrt{2} G'_{AB}{}^{DE}\overline\eta\Gamma_{[C}\psi_{DE]}.
\end{equation}
The last two tems cancel with Eq. (\ref{dog}), leaving a term which matches Eq. (\ref{cfp}). We can
conclude that the boundary condition on $C$ is supersymmetric.

\subsection{Anomaly Cancellation}

Earlier in this section we used the fact that anomaly cancellation requires the combination of
Chern-Simons forms
\begin{equation}
\omega^Y-\frac12\omega^L.\label{comb}
\end{equation}
in the boundary condition for the antisymmetric field.
This combination orginates in the 12-form $I_{12}$ which generates the gauge, gravity and
supergravity anomalies. Horava and Witten obtained an expression for this 12-form by combining
gaugino and gravitino contributions,  
\begin{equation}
I_{12}={1\over 12(2\pi)^5}(I_4^3-4I_4X_8),\label{aform}
\end{equation}
where
\begin{eqnarray}
I_4&=&{\rm tr}\,F^2-\frac12{\rm tr}\, R^2\\
X_8&=&-\frac18{\rm tr}\,R^4+\frac{1}{32}({\rm tr}\, R^2)^2.\label{x8}
\end{eqnarray}
The usual notation convention is used now where exterior products are implied rather than explicit.
The combination (\ref{comb}) allows the gauge variation of the $CGG$ term in the action to cancel
the anomalies descended from $I_4^3$.

We have found that supersymmetry demands $G$-flux terms to appear in addition to the Chern-Simons
terms in the boundary conditions. Anomaly cancellation will only occur if these terms also appear
in $I_4$,
\begin{equation}
I_4={\rm tr}\,F^2-\frac12{\rm tr}\, R^-{}^2+d\omega^G+O(\alpha^6),
\end{equation}
where $\omega^G$ was given in Eq. (\ref{omegagdef}). Note that we can only determine the $G_{NABC}$
terms at this order, even though $G_{ABCD}$ terms may also contribute to $I_4$. As a matter of fact,
both $G$-flux and extrinsic curvature terms can contribute to the anomaly, since these where both
dropped from the original anomaly calculations.

The anomaly 12-form has now been calculated with $G$-flux terms by  Lukic and Moore
\cite{Lukic:2007aj}. Unfortunately, a direct comparison is complicated for a number of reasons.
Firstly,  Lukic and Moore include a `$G\chi\chi$' term in their boundary action as suggested by
Horava and Witten \cite{Horava:1996ma}, but which is not allowed in the improved theory. Secondly,
the fields in the direct anomaly calculation satisfy the background field equations. The boundary
conditions only give the $I_4^3$ part of the anomaly and we need the full expression to compare
when subject to field equations. 

For the remainder of this section we turn from the $G$-flux terms to higher order curvature terms.
The $I_4X_8$ term in $I_{12}$ can be cancelled by a Green-Schwarz term $CX_8$ in the
$11$-dimensional action \cite{Lukas:1998ew}. To make sense of this term, $X_8$ has to be defined in
the 11-dimensional bulk so that it reduces to (\ref{x8}) on the boundary.  There is no need to
modify the boundary condition on the $C$ field on account of the $CX_8$ term, the boundary
condition being determined only by $I_4$ as long as the anomaly takes the general form
(\ref{aform}) so that the Green Schwarz mechanism can be applied. 

We have seen already how anomaly cancellation leads to a unique combination of
curvature-squared terms in the boundary action. This occurs also at higher orders in curvature.
The boundary condition on the $C$ field is determined by $I_4$. The other boundary conditions are
then fixed by supersymmetry. In turn the boundary action, which is determined
by the boundary conditions, must also be fixed by $I_4$. Terms which are higher order than the
square of the curvature can arise in this way from extrinsic curvature contributions to the
anomaly. These can be replaced by higher order intrinsic curvature terms by using the boundary
condition on the extrinsic curvature.

As an example, we could consider the `$\epsilon^2 R^4$' interaction terms in the supergravity action
which are related to the `$CX_8$' term by supersymmetry \cite{Lukas:1998ew}. These will bring in
boundary terms of the form `$\epsilon^2 K R^3$', equivalent to `$\epsilon K^2 R$' after we apply
the boundary condition on the extrinsic curvature $K$. Anomaly cancellation will now require
another modification to $I_4$, introducing `$K^2 R$' terms. In principle, we could reconstruct the
extrinsic curvature terms in the anomaly term this way. Comparing these against a direct
calculation of the anomaly would be a highly non-trivial consistency check.

Another reason this approach may be of interest is that the arguments made so far for the low energy
effective action can be applied equally well to the quantum field theory effective action for
supergravity on a manifold with boundary. By including a boundary we introduce anomalies. If these
can be cancelled
by a Green-Schwarz type of mechanism, then local terms in the boundary action are severely
constrained. In
particular, any supersymmetric counterterms to the theory which required boundary contributions
to the action would not be allowed.

\section{conclusion}

11-dimensional supergravity on a manifold with boundary shows an amazing robustness. 
At each successive order in derivatives, the anomaly-free extension of the theory is very tightly
constrained, but so far this has not forced any internal contradictions. This is consistent  
with the idea that the construction produces the low-energy limit of a well-defined theory of some
kind. 

The results for the higher order terms obtained in this paper can be summarised as follows.  First
of all, the curvature-squared terms in the
boundary of the supergravity action up to fifth order in derivatives are
\begin{equation}
S_{RR}={\epsilon\over 2\kappa_{11}^2}\int_{\cal \partial M}dv
\left(\frac14R^-{}_{ABCD}R^-{}^{ABCD}+
\frac12\bar\psi_{BC}\Gamma^AD_A(\omega)\psi^{BC}
+\frac14\bar\psi_A\Gamma^{BC}\Gamma^AR_{BCDE}\psi^{DE}\right),\label{rract}
\end{equation}
where $\psi_{AB}$ is the gravitino curvature Eq. (\ref{gcdef}) and  the minus
superscript indicates use of the modified Lorentz connection with $G$-flux terms Eq. (\ref{msc}). 
The theory now has vanishing gravity and supergravity anomalies, as well as the vanishing gauge
anomaly which existed previously. The supersymmetric boundary conditions up to fifth order in
derivatives and two fermi fields are,
\begin{eqnarray}
C_{ABC}&=&{\sqrt{2}\over 24}\epsilon
\left(\omega^L_{ABC}+\omega^\psi_{ABC}+\omega^G_{ABC}\right)\\
P_+\psi_A&=&-{1\over 24}\epsilon\left(
\Gamma_A{}^{BC}-10\delta_A{}^B\Gamma^C\right)R^-{}_{BCDE}\psi^{DE}
+\epsilon f_A{}^{BC}\psi_{BC}\\
\hat K_{AB}&=&-\frac12\epsilon\left(R^-{}_A{}{}_{CDE}R^-{}_{B}{}^{CDE}
-{1\over 12}g_{AB}R^-{}_{CDEF}R^-{}^{CDEF}\right)+\hbox{$\psi$-terms}.\label{kabbc}
\end{eqnarray}
where the $\psi$-terms in Eq. (\ref{kabbc}) can be obtained by variation of the full action whilst
keeping the surface connection fixed, and
\begin{eqnarray}
\omega^L_{ABC}&=&\omega^L_{ABC}(\hat\omega^-)\\
\omega^\psi_{ABC}&=&\frac14\overline\psi_{DE} \Gamma_{ABC}\psi^{DE}
-6\overline\psi_{N[A}\psi_{BC]}-
{\sqrt{2}\over 6}G_{NPQR}\overline\psi_{[AB}\Gamma^{PQR}\psi_{C]}\\
\omega^G_{ABC}&=&-\frac{1}{3} *\hat G'_{ABC}{}^{DEF}\hat G_{NDEF}
+3\hat G_{N[A}{}^{DE} \hat G'_{BC]DE}\\
f_A{}^{BC}&=&{\sqrt{2}\over 24}\Gamma^{PQR}\delta_A{}^{[B}
G'{}^{C]}{}_{PQR}+{\sqrt{2}\over 48}\Gamma_A\Gamma^{PQ}
G'_{PQ}{}^{BC}.
\end{eqnarray}
Hats denote the supercovariant quantity constructed by adding fermion terms and prime denotes a
derivative in the normal direction. 

The flux terms imply new contributions to the gravitino anomaly of supergravity on a manifold
with boundary. Some progress has been made in calculating these terms directly \cite{Lukic:2007aj},
but so far more work is needed for a full comparison to be made. A direct calculation of
the gravitational anomaly including  flux terms and extrinsic curvatures would give an important
check that the Green-Schwarz mechanism can be used at higher derivative orders to obtain an
anomaly-free theory.

Introducing the boundary means that total divergences which are usually discarded when discussing
supersymmetry have to be retained. These total divergences are particularly dangerous when they
start to interfere with the anomaly cancellation mechanism, as described in section IIIC.   This
restricts the addition of new bulk interaction terms to the supergravity action
\cite{Green:2005ba}. There may also be important implications for the allowed counterterms in
quantised 11-dimensional supergravity \cite{Bern:2007hh}, and it would be interesting to examine
both interaction and counterterms on a manifold with boundary. 

\bibliography{paper.bib}

\begin{thebibliography}{22}
\expandafter\ifx\csname natexlab\endcsname\relax\def\natexlab#1{#1}\fi
\expandafter\ifx\csname bibnamefont\endcsname\relax
  \def\bibnamefont#1{#1}\fi
\expandafter\ifx\csname bibfnamefont\endcsname\relax
  \def\bibfnamefont#1{#1}\fi
\expandafter\ifx\csname citenamefont\endcsname\relax
  \def\citenamefont#1{#1}\fi
\expandafter\ifx\csname url\endcsname\relax
  \def\url#1{\texttt{#1}}\fi
\expandafter\ifx\csname urlprefix\endcsname\relax\def\urlprefix{URL }\fi
\providecommand{\bibinfo}[2]{#2}
\providecommand{\eprint}[2][]{\url{#2}}

\bibitem[{\citenamefont{Horava and Witten}(1996{\natexlab{a}})}]{Horava:1995qa}
\bibinfo{author}{\bibfnamefont{P.}~\bibnamefont{Horava}} \bibnamefont{and}
  \bibinfo{author}{\bibfnamefont{E.}~\bibnamefont{Witten}},
  \bibinfo{journal}{Nucl. Phys.} \textbf{\bibinfo{volume}{B460}},
  \bibinfo{pages}{506} (\bibinfo{year}{1996}{\natexlab{a}}),
  \eprint{hep-th/9510209}.

\bibitem[{\citenamefont{Horava and Witten}(1996{\natexlab{b}})}]{Horava:1996ma}
\bibinfo{author}{\bibfnamefont{P.}~\bibnamefont{Horava}} \bibnamefont{and}
  \bibinfo{author}{\bibfnamefont{E.}~\bibnamefont{Witten}},
  \bibinfo{journal}{Nucl. Phys.} \textbf{\bibinfo{volume}{B475}},
  \bibinfo{pages}{94} (\bibinfo{year}{1996}{\natexlab{b}}),
  \eprint{hep-th/9603142}.

\bibitem[{\citenamefont{Witten}(1996)}]{Witten:1996mz}
\bibinfo{author}{\bibfnamefont{E.}~\bibnamefont{Witten}},
  \bibinfo{journal}{Nucl. Phys.} \textbf{\bibinfo{volume}{B471}},
  \bibinfo{pages}{135} (\bibinfo{year}{1996}), \eprint{hep-th/9602070}.

\bibitem[{\citenamefont{Banks and Dine}(1996)}]{banks96}
\bibinfo{author}{\bibfnamefont{T.}~\bibnamefont{Banks}} \bibnamefont{and}
  \bibinfo{author}{\bibfnamefont{M.}~\bibnamefont{Dine}},
  \bibinfo{journal}{Nucl. Phys.} \textbf{\bibinfo{volume}{B479}},
  \bibinfo{pages}{173} (\bibinfo{year}{1996}), \eprint{hep-th/9605136}.

\bibitem[{\citenamefont{Moss}(2003)}]{Moss:2003bk}
\bibinfo{author}{\bibfnamefont{I.~G.} \bibnamefont{Moss}},
  \bibinfo{journal}{Phys. Lett.} \textbf{\bibinfo{volume}{B577}},
  \bibinfo{pages}{71} (\bibinfo{year}{2003}), \eprint{hep-th/0308159}.

\bibitem[{\citenamefont{Moss}(2005)}]{Moss:2004ck}
\bibinfo{author}{\bibfnamefont{I.~G.} \bibnamefont{Moss}},
  \bibinfo{journal}{Nucl. Phys.} \textbf{\bibinfo{volume}{B729}},
  \bibinfo{pages}{179} (\bibinfo{year}{2005}), \eprint{hep-th/0403106}.

\bibitem[{\citenamefont{Moss}(2006)}]{Moss:2005zw}
\bibinfo{author}{\bibfnamefont{I.~G.} \bibnamefont{Moss}},
  \bibinfo{journal}{Phys. Lett.} \textbf{\bibinfo{volume}{B637}},
  \bibinfo{pages}{93} (\bibinfo{year}{2006}), \eprint{hep-th/0508227}.

\bibitem[{\citenamefont{Lukas et~al.}(1999{\natexlab{a}})\citenamefont{Lukas,
  Ovrut, and Waldram}}]{Lukas:1998ew}
\bibinfo{author}{\bibfnamefont{A.}~\bibnamefont{Lukas}},
  \bibinfo{author}{\bibfnamefont{B.~A.} \bibnamefont{Ovrut}}, \bibnamefont{and}
  \bibinfo{author}{\bibfnamefont{D.}~\bibnamefont{Waldram}},
  \bibinfo{journal}{Nucl. Phys.} \textbf{\bibinfo{volume}{B540}},
  \bibinfo{pages}{230} (\bibinfo{year}{1999}{\natexlab{a}}),
  \eprint{hep-th/9801087}.

\bibitem[{\citenamefont{Lukas et~al.}(1999{\natexlab{b}})\citenamefont{Lukas,
  Ovrut, Stelle, and Waldram}}]{Lukas:1998tt}
\bibinfo{author}{\bibfnamefont{A.}~\bibnamefont{Lukas}},
  \bibinfo{author}{\bibfnamefont{B.~A.} \bibnamefont{Ovrut}},
  \bibinfo{author}{\bibfnamefont{K.~S.} \bibnamefont{Stelle}},
  \bibnamefont{and} \bibinfo{author}{\bibfnamefont{D.}~\bibnamefont{Waldram}},
  \bibinfo{journal}{Nucl. Phys.} \textbf{\bibinfo{volume}{B552}},
  \bibinfo{pages}{246} (\bibinfo{year}{1999}{\natexlab{b}}),
  \eprint{hep-th/9806051}.

\bibitem[{\citenamefont{Buchbinder and Ovrut}(2004)}]{Buchbinder:2003pi}
\bibinfo{author}{\bibfnamefont{E.~I.} \bibnamefont{Buchbinder}}
  \bibnamefont{and} \bibinfo{author}{\bibfnamefont{B.~A.} \bibnamefont{Ovrut}},
  \bibinfo{journal}{Phys. Rev.} \textbf{\bibinfo{volume}{D69}},
  \bibinfo{pages}{086010} (\bibinfo{year}{2004}), \eprint{hep-th/0310112}.

\bibitem[{\citenamefont{Braun and Ovrut}(2006)}]{Braun:2006th}
\bibinfo{author}{\bibfnamefont{V.}~\bibnamefont{Braun}} \bibnamefont{and}
  \bibinfo{author}{\bibfnamefont{B.~A.} \bibnamefont{Ovrut}},
  \bibinfo{journal}{JHEP} \textbf{\bibinfo{volume}{07}}, \bibinfo{pages}{035}
  (\bibinfo{year}{2006}), \eprint{hep-th/0603088}.

\bibitem[{\citenamefont{Gray et~al.}(2007)\citenamefont{Gray, Lukas, and
  Ovrut}}]{Gray:2007qy}
\bibinfo{author}{\bibfnamefont{J.}~\bibnamefont{Gray}},
  \bibinfo{author}{\bibfnamefont{A.}~\bibnamefont{Lukas}}, \bibnamefont{and}
  \bibinfo{author}{\bibfnamefont{B.}~\bibnamefont{Ovrut}},
  \bibinfo{journal}{Phys. Rev.} \textbf{\bibinfo{volume}{D76}},
  \bibinfo{pages}{126012} (\bibinfo{year}{2007}), \eprint{0709.2914}.

\bibitem[{\citenamefont{Green and Schwarz}(1984)}]{Green:1984sg}
\bibinfo{author}{\bibfnamefont{M.~B.} \bibnamefont{Green}} \bibnamefont{and}
  \bibinfo{author}{\bibfnamefont{J.~H.} \bibnamefont{Schwarz}},
  \bibinfo{journal}{Phys. Lett.} \textbf{\bibinfo{volume}{B149}},
  \bibinfo{pages}{117} (\bibinfo{year}{1984}).

\bibitem[{\citenamefont{Cremmer et~al.}(1978)\citenamefont{Cremmer, Julia, and
  Scherk}}]{cremmer78}
\bibinfo{author}{\bibfnamefont{E.}~\bibnamefont{Cremmer}},
  \bibinfo{author}{\bibfnamefont{B.}~\bibnamefont{Julia}}, \bibnamefont{and}
  \bibinfo{author}{\bibfnamefont{J.}~\bibnamefont{Scherk}},
  \bibinfo{journal}{Phys Lett} \textbf{\bibinfo{volume}{B76}},
  \bibinfo{pages}{409} (\bibinfo{year}{1978}).

\bibitem[{\citenamefont{Luckock and Moss}(1989)}]{luckock89}
\bibinfo{author}{\bibfnamefont{H.~C.} \bibnamefont{Luckock}} \bibnamefont{and}
  \bibinfo{author}{\bibfnamefont{I.~G.} \bibnamefont{Moss}},
  \bibinfo{journal}{Class Quantum grav} \textbf{\bibinfo{volume}{6}},
  \bibinfo{pages}{1993} (\bibinfo{year}{1989}).

\bibitem[{\citenamefont{Weinberg}(2002)}]{Weinberg:2000cr}
\bibinfo{author}{\bibfnamefont{S.}~\bibnamefont{Weinberg}},
  \emph{\bibinfo{title}{{The quantum theory of fields. Vol. 3: Supersymmetry}}}
  (\bibinfo{publisher}{Cambridge, UK: Univ. Pr. (2000) 419 p},
  \bibinfo{year}{2002}).

\bibitem[{\citenamefont{Bergshoeff and
  de~Roo}(1989{\natexlab{a}})}]{Bergshoeff:1989de}
\bibinfo{author}{\bibfnamefont{E.~A.} \bibnamefont{Bergshoeff}}
  \bibnamefont{and} \bibinfo{author}{\bibfnamefont{M.}~\bibnamefont{de~Roo}},
  \bibinfo{journal}{Nucl. Phys.} \textbf{\bibinfo{volume}{B328}},
  \bibinfo{pages}{439} (\bibinfo{year}{1989}{\natexlab{a}}).

\bibitem[{\citenamefont{Bergshoeff and
  de~Roo}(1989{\natexlab{b}})}]{Bergshoeff:1988nn}
\bibinfo{author}{\bibfnamefont{E.}~\bibnamefont{Bergshoeff}} \bibnamefont{and}
  \bibinfo{author}{\bibfnamefont{M.}~\bibnamefont{de~Roo}},
  \bibinfo{journal}{Phys. Lett.} \textbf{\bibinfo{volume}{B218}},
  \bibinfo{pages}{210} (\bibinfo{year}{1989}{\natexlab{b}}).

\bibitem[{\citenamefont{Gross and Sloan}(1987)}]{Gross:1986mw}
\bibinfo{author}{\bibfnamefont{D.~J.} \bibnamefont{Gross}} \bibnamefont{and}
  \bibinfo{author}{\bibfnamefont{J.~H.} \bibnamefont{Sloan}},
  \bibinfo{journal}{Nucl. Phys.} \textbf{\bibinfo{volume}{B291}},
  \bibinfo{pages}{41} (\bibinfo{year}{1987}).

\bibitem[{\citenamefont{Lukic and Moore}(2007)}]{Lukic:2007aj}
\bibinfo{author}{\bibfnamefont{S.}~\bibnamefont{Lukic}} \bibnamefont{and}
  \bibinfo{author}{\bibfnamefont{G.~W.} \bibnamefont{Moore}}
  (\bibinfo{year}{2007}), \eprint{hep-th/0702160}.

\bibitem[{\citenamefont{Green and Vanhove}(2006)}]{Green:2005ba}
\bibinfo{author}{\bibfnamefont{M.~B.} \bibnamefont{Green}} \bibnamefont{and}
  \bibinfo{author}{\bibfnamefont{P.}~\bibnamefont{Vanhove}},
  \bibinfo{journal}{JHEP} \textbf{\bibinfo{volume}{01}}, \bibinfo{pages}{093}
  (\bibinfo{year}{2006}), \eprint{hep-th/0510027}.

\bibitem[{\citenamefont{Bern et~al.}(2007)}]{Bern:2007hh}
\bibinfo{author}{\bibfnamefont{Z.}~\bibnamefont{Bern}} \bibnamefont{et~al.},
  \bibinfo{journal}{Phys. Rev. Lett.} \textbf{\bibinfo{volume}{98}},
  \bibinfo{pages}{161303} (\bibinfo{year}{2007}), \eprint{hep-th/0702112}.

\end{thebibliography}

\end{document}